\documentclass[aps,prl,preprint,superscriptaddress]{revtex4-1}

\usepackage{graphicx}
\usepackage{amsmath}

\begin{document}

\title{Polaronic behavior in a weak coupling superconductor}

\author{Adrian G Swartz}
\thanks{Contributed equally}
\affiliation{Department of Applied Physics, Stanford University, Stanford, California 94305, USA}
\affiliation{Stanford Institute for Materials and Energy Sciences, SLAC National Accelerator Laboratory \& Stanford University, Menlo Park, California 94025, USA}
\affiliation{Geballe Laboratory for Advanced Materials, Stanford University, Stanford, California 94305, USA}
\email{aswartz@stanford.edu}

\author{Hisashi Inoue}
\thanks{Contributed equally}
\affiliation{Department of Applied Physics, Stanford University, Stanford, California 94305, USA}
\affiliation{Stanford Institute for Materials and Energy Sciences, SLAC National Accelerator Laboratory \& Stanford University, Menlo Park, California 94025, USA}
\affiliation{Geballe Laboratory for Advanced Materials, Stanford University, Stanford, California 94305, USA}

\author{Tyler A Merz}
\affiliation{Department of Applied Physics, Stanford University, Stanford, California 94305, USA}
\affiliation{Geballe Laboratory for Advanced Materials, Stanford University, Stanford, California 94305, USA}

\author{Yasuyuki Hikita}
\affiliation{Stanford Institute for Materials and Energy Sciences, SLAC National Accelerator Laboratory \& Stanford University, Menlo Park, California 94025, USA}

\author{Srinivas Raghu}
\affiliation{Stanford Institute for Materials and Energy Sciences, SLAC National Accelerator Laboratory \& Stanford University, Menlo Park, California 94025, USA}
\affiliation{Department of Physics, Stanford University, Stanford, California 94305, USA}

\author{Thomas P. Devereaux}
\affiliation{Stanford Institute for Materials and Energy Sciences, SLAC National Accelerator Laboratory \& Stanford University, Menlo Park, California 94025, USA}
\affiliation{Geballe Laboratory for Advanced Materials, Stanford University, Stanford, California 94305, USA}

\author{Steven Johnston}
\affiliation{Department of Physics and Astronomy, University of Tennessee, Knoxville, Tennessee 37996, USA}

\author{Harold Y Hwang}
\affiliation{Department of Applied Physics, Stanford University, Stanford, California 94305, USA}
\affiliation{Stanford Institute for Materials and Energy Sciences, SLAC National Accelerator Laboratory \& Stanford University, Menlo Park, California 94025, USA}
\affiliation{Geballe Laboratory for Advanced Materials, Stanford University, Stanford, California 94305, USA}

\begin{abstract}
The nature of superconductivity in the dilute semiconductor SrTiO$_3$ has remained an open question for more than 50 years. 
The extremely low carrier densities ($10^{18}$ - $10^{20}$ cm$^{-3}$) at which superconductivity occurs suggests an unconventional origin of superconductivity outside of the adiabatic limit on which the Bardeen-Cooper-Schrieffer (BCS) and Migdal-Eliashberg (ME) theories are based. 
We take advantage of a newly developed method for engineering band alignments at oxide interfaces and access the electronic structure of Nb-doped SrTiO$_3$ using high resolution tunneling spectroscopy. We observe strong coupling to the highest energy longitudinal optic (LO) phonon branch and estimate the doping evolution of the dimensionless electron-phonon interaction strength ($\lambda$). 
Upon cooling below the superconducting transition temperature ($T_{\mathrm{c}}$), we observe a single superconducting gap corresponding to the weak-coupling limit of BCS theory, indicating an order of magnitude smaller coupling ($\lambda_{\textrm{BCS}} \approx 0.1$). 
These results suggest that despite the strong normal state interaction with electrons, the highest LO phonon does not provide a dominant contribution to pairing.
They further demonstrate that SrTiO$_3$ is an ideal system to probe superconductivity over a wide range of carrier density, adiabatic parameter, and electron-phonon coupling strength.
\end{abstract}

\maketitle

Electron doped SrTiO$_3$ undergoes a superconducting transition at low temperature following a dome-like behavior that spans the largest range of carrier density ($n$) and Fermi energy ($E_F$) of any material \cite{Schooley:1964, Koonce:1967, Lin:2014}.
For much of the superconducting dome, the typical phonon energies ($\hbar\omega_{ph}$) are comparable to or larger than $E_F$, yielding an adiabatic ratio of $\hbar\omega_{ph}/E_F>1$. 
The nature of superconductivity in the limit of so few electrons is a long-standing problem in condensed matter physics. 
Several theoretical approaches consider pairing due to electron-phonon ($e$-ph) coupling \cite{Appel:1969,Takada:1980,Ruhman:2016,Rosenstein:2016,Klimin:2017}. 
In parallel, there is experimental evidence for the formation of large polarons -- quasiparticles arising from strong $e$-ph coupling \cite{Ahrens:2007,Van:2008,Chang:2010,Mazin:2011,Chen:2015,Wang:2016a}.
Pairing scenarios involving plasmons or quantum critical ferroelectric fluctuations have also been proposed \cite{Takada:1980,Ruhman:2016,Edge:2015a,Rischau:2017}. 
These fundamental issues are also relevant for emerging two-dimensional (2D) superconductors such as LaAlO$_3$/SrTiO$_3$ \cite{Caviglia:2008}, $\delta$-doped SrTiO$_3$ \cite{Kozuka:2009}, and FeSe/SrTiO$_3$ \cite{Lee:2014}.

Tunneling spectroscopy, which we employ here, has been instrumental in studying superconductivity, $e$-ph coupling, and their interrelationship \cite{Carbotte:1990}.
The onset of superconductivity creates a gap ($\Delta$) in the density of states (DOS), which can be resolved in high-resolution tunneling experiments. 
In addition, if the coupling is strong, the $e$-ph interaction can induce measurable renormalizations to the electronic band structure \cite{McMillan:1965,Lanzara:2001}.
Recently, there have been compelling angle-resolved photoemission spectroscopy (ARPES) reports of band renormalization at the surface of SrTiO$_3$, where replica or shake-off states arising from polaronic coupling to the highest energy LO phonon \cite{Chen:2015,Wang:2016a} are observed. 
Such dramatic modifications of the electronic structure should also be directly observable in tunneling experiments.
Moreover, while ARPES experiments are limited to temperature regimes above $T_c$, tunneling spectroscopy offers a unique opportunity to examine the relationship between the superconducting electronic structure and $e$-ph coupling in the same sample.

Energetically resolved tunneling is a challenge in bulk SrTiO$_3$ due to the large dielectric constant ($\varepsilon_0$ $\approx$ 20,000 at low temperatures), which generates long depletion lengths ($>$ 100 nm). 
Early experiments attempted to examine the superconducting gap using Schottky contacts to heavily-doped SrTiO$_3$ and tunnel through the depletion region; they indicated coupling to the LO phonon modes and multiple superconducting gaps, stimulating considerations of multiband superconductivity \cite{Sroubek:1969, Binnig:1980, Fernandes:2013}.
However, because of the uncontrolled and varying interfacial carrier densities and depletion lengths, experiments probing the doping evolution of the bulk material were unfeasible. 
Recently, pioneering tunneling experiments have examined the 2D superconducting layer formed at the interface between LaAlO$_3$ and undoped SrTiO$_3$ (LaAlO$_3$/SrTiO$_3$) \cite{Richter:2013,Boschker:2015}, utilizing the thin LaAlO$_3$ as the tunnel barrier.
Similar to bulk SrTiO$_3$, LaAlO$_3$/SrTiO$_3$ exhibits a superconducting transition below 0.4 K \cite{Reyren:2007}, but in the 2D limit. 
This system exhibits a superconducting dome that can be tuned by a backgate, where the dominant effect of gating is the variation of the mobility edge rather than modulation of the sheet carrier density \cite{Caviglia:2008, Bell:2009}. Moreover, this 2D superconducting ground state is highly complex, as demonstrated by the coexistence of magnetism \cite{Bert:2011}, a high density of both itinerant and localized states, observation of a pseudo-gap \cite{Richter:2013}, and the occupation of multiple subbands \cite{Chen:2016}.  

Here, we demonstrate access to the bulk Nb-doped SrTiO$_3$ electronic structure in $both$ the normal and superconducting state, at carrier densities across the superconducting dome. Using atomically engineered interface tunnel barriers \cite{Yajima:2015, Inoue:2015}, in which ultrathin epitaxial (001) LaAlO$_3$ serves as a polar tunnel barrier in planar tunnel junctions (Fig. 1A,B), we suppress the interfacial depletion layer and ensure that the tunneling spectra reflect the intrinsic bulk properties.
Importantly, in contrast to LaAlO$_3$/SrTiO$_3$, when ultrathin epitaxial layers of LaAlO$_3$ are placed on top of metallic Nb-doped SrTiO$_3$, there is no additional electronic reconstruction induced at the interface for the nominal doping concentrations studied here.
This is consistent with the observation of the crossover from excess interface charge induced by a polar discontinuity, to metallic screening across a metal-insulator transition \cite{Mundy:2014}.
Instead, the LaAlO$_3$ provides an interfacial dipole, shifting the band alignments to remove the depletion region (Fig. 1C,D).
This simple technique is analogous to band alignment manipulations in semiconductor heterostructures \cite{Tung:2014}, and has been demonstrated to be highly effective across several oxide systems and applications \cite{Inoue:2015,Tachikawa:2015,Hikita:2016}.

%
%
\begin{figure}[t]
\includegraphics[width=85mm]{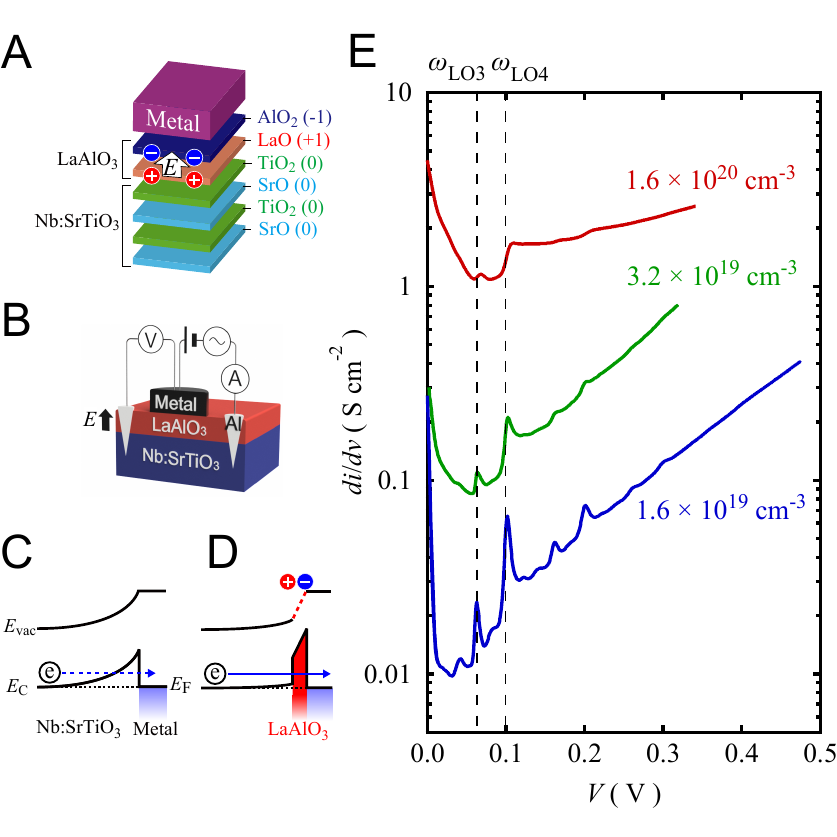}
\caption{\label{fig1} 
Dipole controlled interfaces for high resolution tunneling spectroscopy of Nb:SrTiO$_3$. Schematic diagrams of the tunneling junction depicting: A, the atomic layer stacking with an interfacial LaAlO$_3$ dipole layer which serves as the tunneling barrier; B, device measurement geometry; C-D, resulting band diagrams without and with the insertion of the interface dipole. E, Experimental conductance ($di/dv$) spectra in the normal state for three characteristic doping concentrations measured at $T=2$ K.  
}
\end{figure}

\section*{Results}
\subsection*{Tunneling spectroscopy in the normal state}
Figure 1E plots the measured differential conductance ($\sigma = di/dv$) in the positive bias regime (electron extraction from SrTiO$_3$) for several characteristic doping concentrations, with $E_F \approx$ 13, 23, and 61 meV \cite{supplement}.
Complete details on the sample preparation and measurement are provided in the methods and supplemental information \cite{supplement}.
The  asymmetry in bias polarity is due to the highly unequal DOS of the semiconducting and metallic electrodes, producing a global minimum set by the SrTiO$_3$ Fermi level. 
Additionally, we observe clear features of enhanced conductance at 36.0, 60.5, and 98.8 meV, corresponding with the energies of bulk SrTiO$_3$ polar LO2, LO3, and LO4 phonon modes, respectively, and not barrier LaAlO$_3$ modes \cite{supplement,Petzelt:2001, Vogt:1981}.
To examine these features with higher sensitivity, we directly measure $d^2i$/$dv^2$ as shown in Fig. 2A.  
The data clearly indicate phonon interactions at energies above the highest energy LO phonon branch ($eV > 0.1$ V). 
For all doping concentrations examined here, the dominant spectral features appear at regular intervals of energy $\hbar\omega_{\textrm{LO4}}$. 
In the low density limit ($E_F$ = 13 meV), additional structure is found up to fourth order in LO4 (i.e. $2\hbar\omega_{\textrm{LO4}}$, $3\hbar\omega_{\textrm{LO4}}$, and $4\hbar\omega_{\textrm{LO4}}$).  
At higher density ($E_F$ = 61 meV), the higher-order structure arising from LO4 is suppressed, indicating that the $e$-ph coupling is larger at low doping. 
Qualitatively, the observation of interactions involving up to four phonon processes indicates strong $e$-ph coupling.  
For instance, the second-order acoustic phonon structure has been observed for Pb ($\lambda$ = 1.3) \cite{Parks:1969, Schackert:2015}.

\begin{figure}[b]
\includegraphics[width=85mm]{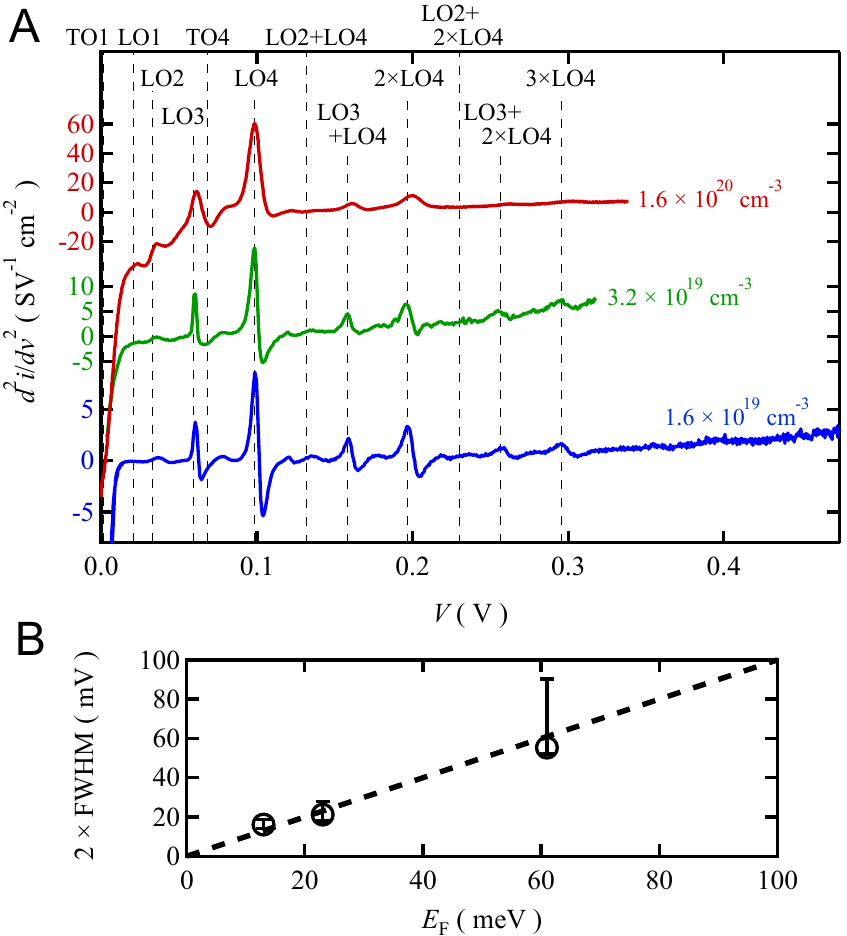}
\caption{\label{fig2} 
Second-harmonic tunneling spectroscopy of $e$-ph coupling in Nb:SrTiO$_3$. A, Measured $d^2i/dv^2$ raw data ($T = 2$ K) for three samples with the following doping concentrations: 1.6 $\times$ 10$^{19}$ cm$^{-3}$, 3.2 $\times$ 10$^{19}$ cm$^{-3}$, and 1.6 $\times$ 10$^{20}$ cm$^{-3}$. Dashed lines indicate the transverse optic (TO) and LO phonon modes from infrared reflectivity and Raman spectroscopy on bulk crystals (Refs. 39 and 40), as well as some of their higher harmonics.
B, Extracted LO4 $e$-ph linewidth (defined as twice the full width at half maximum (FWHM) of the LO4 feature in $di/dv$ (38)) vs. bulk $E_F$. The lineshape evolves as $\hbar \omega_{\textrm{LO4}}/E_F \rightarrow 1$, increasing the uncertainty towards larger linewidth values (38).  
}
\end{figure}

In general, tunneling spectroscopy can exhibit features arising from both elastic (DOS renormalizations via virtual phonons) and inelastic (real phonon emission) processes \cite{Wolf:2012, Adkins:1985, McMillan:1965, Conley:1967, Jaklevic:1966}.
Parsing the contributions is a subtle issue and depends upon on the material system, quality of the junction, and experimental probe (i.e. scanning tunneling microscopy, point contact, or planar junction). Here, the observation of peaks, rather than steps in the conductance, rules out threshold inelastic processes \cite{Wolf:2012, Adkins:1985}. 
Further, the lineshape of the $e$-ph spectral features seen in Figs. 1E and 2A evolve monotonically as a function of doping, depending upon the ratio $ \hbar\omega_\textrm{{LO4}} / E_F $. 
One of the key findings here is the observation of a quantitative correspondence between the LO4 phonon spectral linewidth and the bulk Fermi level, as shown in Fig. 2B. This result demonstrates that the $e$-ph linewidth provides an independent spectroscopic measure of the interfacial $E_F$. Further, Fig. 2B indicates that the interface reflects the intended bulk carrier density and is incompatible with surface accumulation layers.
The quantitative agreement with SrTiO$_3$ bulk phonons, sharp bandwidth-limited ($E_F$) spectra at low densities, and evolution as a function of doping indicate the importance of the $e$-ph coupling within SrTiO$_3$. While we cannot rule out the presence of inelastic processes, as a first attempt to estimate the $e$-ph coupling, we consider that similar LO4 features have been observed by ARPES \cite{Chen:2015,Wang:2016a}, and examine this problem in terms of the many-body interactions within SrTiO$_3$.

%
%
\subsection*{Modeling the electron-phonon coupling in the normal state}
To elucidate the impact that the $e$-ph coupling can have on the electronic bands and  tunneling spectra, we calculate the normal state spectral function for non-interacting electrons in a single parabolic band coupled to an Einstein mode of $\hbar \omega_{ph}$ = 100 meV.  
LO modes in ionic crystals generate long range dipole fields which can interact strongly with charged carriers, leading to an interaction with matrix element $M(\boldsymbol{q}) \propto 1/q$ that is highly momentum dependent \cite{Giustino:2017}.
In the real system, the coupling is screened such that $|M({\bf q})|^2 \rightarrow |M({\bf q})|^2/\epsilon({\bf q},\omega)$, where $\epsilon({\bf q},\omega)$ is the dielectric function arising from all sources. 
To calculate the electronic self-energy in the normal state, we consider dynamic screening in the long-wavelength limit such that $\epsilon({\bf q} = 0,\omega)\approx 1 - \Omega_{pl}^2 / \omega_{ph}^2$, where $\Omega_{pl}$ is the plasma frequency. Therefore, $|M({\bf q})|^2 = M_0^2/q^2$, where $M_0$ contains all of the momentum independent factors and is adjusted to set the value of $\lambda$. Full details of the self-energy calculations are provided in the supplemental information \cite{supplement}.

\begin{figure}[t]
\includegraphics[width=85mm]{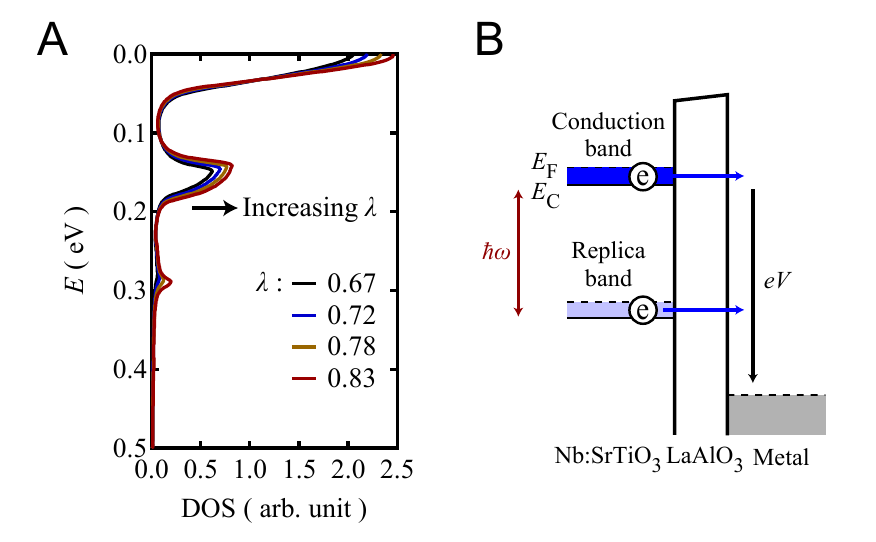}
\caption{\label{fig3} 
Renormalization of the electron band structure due to a polar $e$-ph interaction. A, Calculated density of states at several values of $\lambda$ for a single parabolic band ($E_F = 40$ meV and $m^* = 1.79m_0$, where $m^*$ is the effective mass and $m_0$ is the bare electron mass) coupled to a single dispersionless phonon mode ($\hbar \omega_{ph} = 100$ meV). The momentum dependence of the polar interaction generates replicas of the main electronic band, which are offset from the main band by multiples of the phonon energy. Further details are provided in the Supplemental Information (38). B, Schematic depicting tunneling including the $e$-ph induced structure in the density of states.  
}
\end{figure}

Figure 3A shows the calculated DOS at several values of coupling strength, indicating renormalized spectral weight appearing at intervals of $\hbar \omega_{ph}$. Second-order phonon structure appears in the calculated self-energy for $\lambda$ > 0.5. The formation of the replicated DOS peaks is primarily due to the forward focusing nature of the polar coupling and is distinct from $q$-linear coupling to acoustic modes or momentum independent Holstein-like couplings \cite{Carbotte:1990,supplement,Lee:2014}. 
As shown conceptually in Fig. 3B in the anti-adiabatic limit ($E_F < \hbar\omega_{ph}$), the replicated DOS with narrow bandwidth generates peaks in the differential conductance as the phonon modes are traversed as a function of bias, enabling a bandwidth-limited resolution of the phonon contributions consistent with the experimental observations. 

We estimate the strength of the $e$-ph interaction ($\lambda$) by comparing the {\it relative} multi-phonon intensities of the measured tunneling spectra with the self-energy calculations.  
This approach relies on the appearance of multi-phonon modes and is distinct from the McMillan and Rowell ``inversion'' method only applicable for $E_F > \hbar\omega_{ph}$ and below the superconducting transition temperature \cite{McMillan:1965,supplement,Boschker:2015}. 
Figure 4A plots the measured $d^2i$/$dv^2$ for $E_F$ = 13 meV (same as Fig. 2A bottom curve), where the largest number of phonon replicas are observed. 
Due to the narrow bandwidth, modes are visible at $m \hbar\omega_{\textrm{LO4}}$ and $\hbar\omega_{\textrm{LO3}} + m \hbar\omega_{\textrm{LO4}}$, where $m$ is an integer (Fig. 4B). 
Here we focus solely on estimating $\lambda$ for LO4 which dominates the tunneling spectra at all dopings.  
A relative increase in the spectral weight of multi-phonon processes is observed with decreasing carrier density (Fig. 4C). 
Figure 4D shows the measured intensity ratios, $I_2/I_1$, for the LO4 mode together with the DOS calculations of Fig. 3. 
We have restricted the self-energy calculation to $\lambda <  1$ where a perturbative expansion is a more controlled approximation \cite{supplement}. 
In the absence of an exact strong coupling theory, an estimate for the $e$-ph coupling can be obtained by extrapolating to larger $\lambda$ values. 
Estimated in this way, $\lambda$ ranges from 0.9 to 1.4, where the highest doping concentration ($n = 1.6\times10^{20}$ cm$^{-3}$) exhibits $\lambda \approx 0.9$, similar to other experiments at the same carrier density \cite{Meevasana:2010,Van:2008}. 
We conclude that more dilute samples exhibit stronger coupling to LO4. 

\begin{figure*}[t]
\centering
\includegraphics[width=17cm]{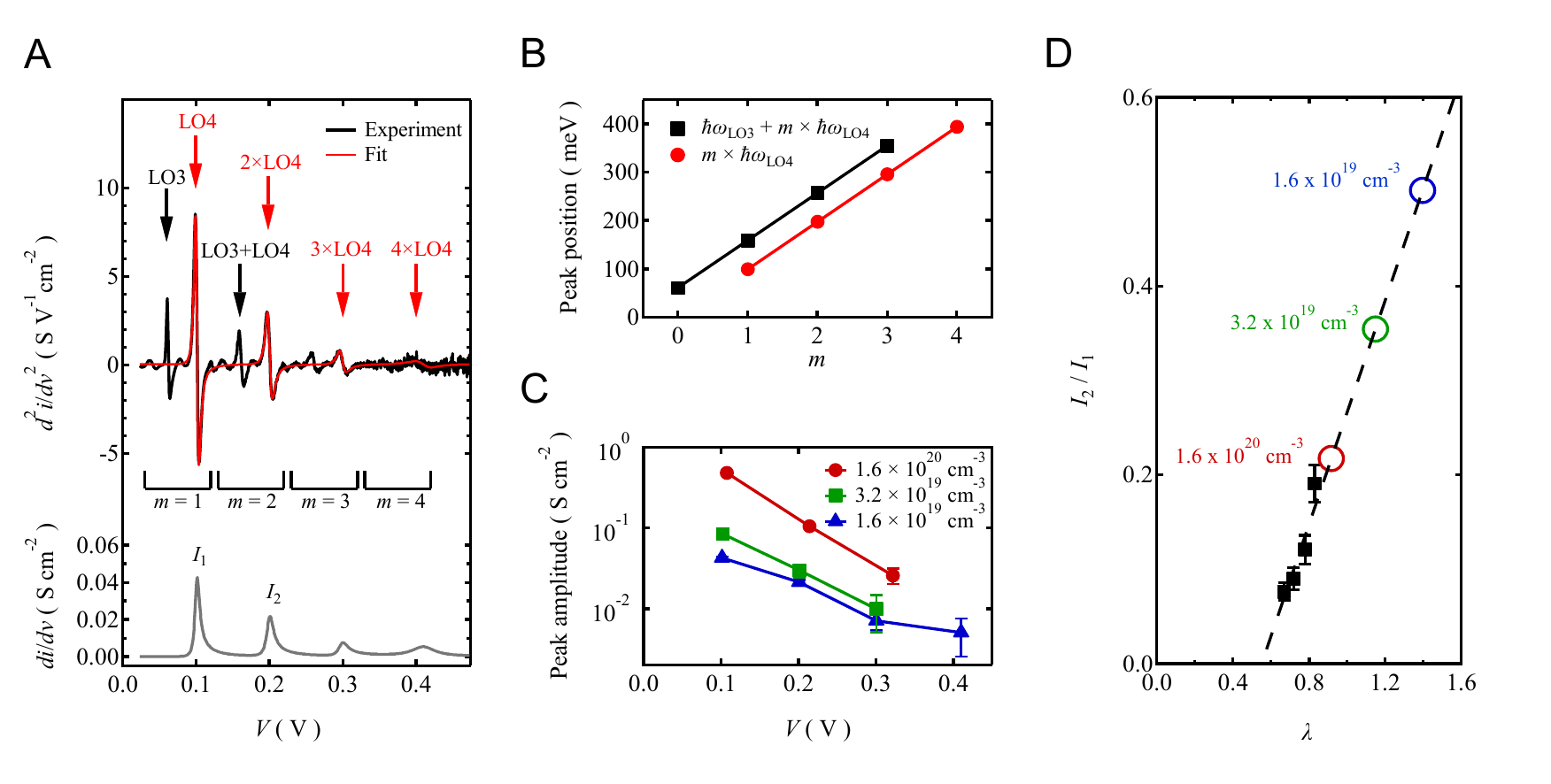}
\caption{\label{fig4} 
Doping dependence of the electron-phonon coupling in Nb:SrTiO$_3$. A, Second harmonic $d^2i/dv^2$ spectra for $n$$\,=\,$$1.6\times10^{19}$ cm$^{-3}$ measured at $T$$\,=\,$2 K (black line, top panel). A smoothly varying background has been subtracted from the raw data (Fig. 2A bottom curve) by a spline fit. The contribution from interactions corresponding to $m\hbar\omega_\textrm{{LO4}}$ ($m$ is an integer) are fit with the derivative of an asymmetric Lorentzian (red line, top panel) for which the integrated $e$-ph contribution to $di/dv$ is shown in the bottom panel (grey curve). B, Extracted energies for multi-phonon processes identified as $\hbar\omega_\textrm{{LO3}} + m\hbar\omega_\textrm{{LO4}}$ and $m\hbar\omega_\textrm{{LO4}}$, demonstrating that LO3 exhibits repeat structure at intervals of LO4.
C, Amplitude of the phonon contribution (LO4 only) as a function of bias for three different doping concentrations. D, Black squares show the intensity ratios for one phonon ($I_1$) and two phonon ($I_2$) processes calculated from the renormalized DOS (Fig. 3A). We linearly extrapolate out of the perturbative regime of $e$-ph coupling strength in the self-energy calculations (black dashed line) and plot the experimental $I_2$/$I_1$ ratios (open circles) to obtain $\lambda$.
}
\end{figure*}

The tunneling results presented thus far clearly indicate that the LO modes significantly modify the electronic properties in SrTiO$_3$ over a wide range of carrier density. This conclusion is consistent with a growing body of experimental evidence for polaron formation in this system. Our results are strikingly similar to recent photoemission measurements of the electronic spectral function $A(\textbf{k},\omega)$ at the surface of SrTiO$_3$, where replica bands were observed at regular intervals of the LO4 phonon mode \cite{Chen:2015,Wang:2016a}. 
In particular, the ratio of the photoemission replica intensities (analogous to $I_2$/$I_1$) are remarkably close to those found here, indicating comparable regimes for the $e$-ph coupling strength. 
These observations are also consistent with magnetotransport, optical conductivity, and heat capacity experiments indicating large polaron formation in SrTiO$_3$ \cite{Ahrens:2007,Van:2008,Mazin:2011}.
Despite the strong $e$-ph coupling, the momentum dependence of the interaction produces relatively modest mass renormalizations, and rather than self-trapped small polarons, highly mobile carriers are found even in the extremely dilute limit \cite{Lin:2014,Kozuka:2008}.

%
%

\subsection*{Superconducting tunneling spectroscopy}
Figure 5A shows a wide-scan $di$/$dv$ spectrum measured below $T_\mathrm{c}$ for $n$ = 2.5$\times$10$^{20}$ cm$^{-3}$ ($E_F$ $\approx$ 70 meV), corresponding to the overdoped side of the Nb:SrTiO$_3$ superconducting dome. Conductivity steps due to $e$-ph interactions are clearly evident, as well as the superconducting gap at low bias. A high-resolution scan (Fig. 5B) shows a single gap which is well fit by the BCS gap function, 
$\sigma_S / \sigma_N = \int_{-\infty}^\infty \nu(E)\frac{\partial f(E+eV)}{\partial E} dE$, 
where $f(E)$ is the Fermi function and 
$\nu(E) = \textrm{Re}[\frac{E-i\Gamma}{\sqrt{(E-i\Gamma)^2-\Delta^2}}]$ 
is the Bogoliubov quasiparticle DOS with a quasiparticle broadening factor $\Gamma$. 
Best fits to the lowest temperature data yield $\Delta$ = 41.9 $\pm$ 0.1 $\mu$eV and $\Gamma$ = 2.2 $\pm$ 0.3 $\mu$eV \cite{Dynes:1978}. The $>90 \%$ suppression of the DOS in the superconducting gap and the low broadening indicates the very high quality of the junction.
For all samples measured we have observed only a single gap, indicating that while at least two $d$-orbital bands are populated for these carrier densities, they are not spectroscopically distinct. 

\begin{figure*}[t]
\centering
\includegraphics[width=17cm]{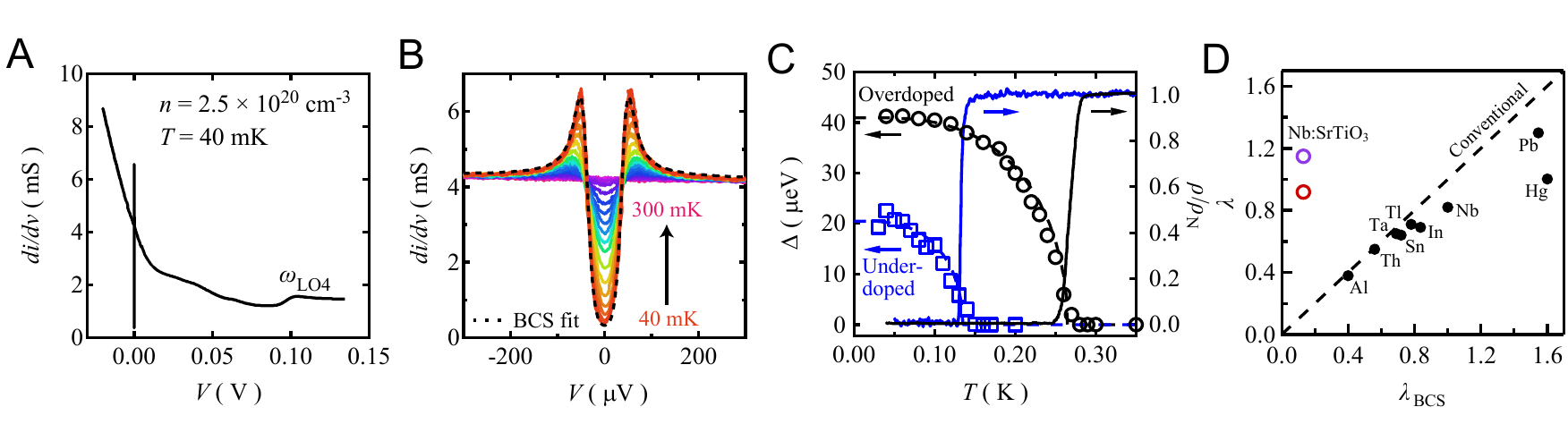}
\caption{\label{fig5} 
Superconducting tunneling spectroscopy of Nb-doped SrTiO$_3$. A, Wide scan $di/dv$ measured at $T = 40$ mK showing the interaction with the high energy phonon modes, as well as the superconducting gap at low bias (near $V$$\,=\,$$0$ V). B, High resolution measurements of the superconducting gap for several temperatures. Dashed curve is the best fit for the 40 mK data to the BCS gap equation. 
For A and B, $n$$\,=\,$$2.5\times10^{20}$ cm$^{-3}$, corresponding with the overdoped side of the Nb:SrTiO$_3$ superconducting dome. 
C, Temperature dependence of the superconducting gap ($\Delta$) (left axis) and resistivity ($\rho$) (right axis), normalized to the normal state value ($\rho_N$) measured at $T$$\,=\,$$400$ mK. Overdoped: $n$$\,=\,$$2.5\times10^{20}$ cm$^{-3}$ (black data), underdoped: $n$$\,=\,$$3.0\times10^{19}$ cm$^{-3}$ (blue data). D, Comparison of the electron-phonon coupling ($\lambda$) estimated from the normal-state spectra with the extracted superconducting pairing strength ($\lambda_{\textrm{BCS}}$) from the superconducting gap. This unconventional regime is evident when compared with conventional superconductors over a broad range of $e$-ph coupling (data compiled from Refs. 20, 21, 42, and 52).
}
\end{figure*}

For conventional phonon-mediated superconductors, the strength of the pairing interaction can be determined by deviations from the universal thermodynamic relations of BCS theory. 
In particular, weak-coupling BCS predicts an exact ratio between the zero temperature gap ($\Delta(T=0) \equiv \Delta_0$) and $T_{\mathrm{c}}$ of $2\Delta_0/k_BT_\mathrm{c} = 3.53$, where $k_B$ is Boltzmann's constant. 
In contrast, strong-coupling superconductors deviate from this value (i.e. Pb with $2\Delta_0/k_BT_\mathrm{c}$ = 4.5 and $\lambda$ = 1.3), which is accounted for by including retardation corrections within ME \cite{Carbotte:1990,Schackert:2015}. 
Due to the signatures of strong $e$-ph coupling observed in the normal state, one might expect SrTiO$_3$ to exhibit significant departures from weak-coupling BCS theory, particularly in the underdoped region where large phonon renormalizations are found. This, however, is not what we observe. 
Figure 5C plots $\Delta(T)$ (open circles, left axis) and the normalized resistivity ($\rho(T)/\rho_N$) (solid lines, right axis) for two characteristic samples corresponding to the overdoped and underdoped side of the Nb-doped superconducting dome.  
The temperature dependence is accurately described by $\Delta(T) = \Delta_0 \tanh{[\frac{\pi}{1.76}\sqrt{\frac{2}{3}1.43(\frac{T_\mathrm{c}}{T}-1)}]}$ \cite{Devereaux:1995}.
We find that the gap closes ($\Delta \rightarrow 0$) with the onset of the resistive $T_\mathrm{c}$, indicating the absence of a pseudogap in the bulk limit. 
Surprisingly, we find $2\Delta_0/k_BT_\mathrm{c} = 3.56$$\,\pm\,$$.03$ and $3.59$$\,\pm\,$$.05$ for the overdoped and underdoped samples, respectively. These values are remarkably close to the universal weak-coupling limit predicted by BCS theory, even compared to the vast majority of conventional superconductors \cite{Carbotte:1990}.  
Such precise agreement is a surprise for a system that clearly violates key assumptions upon which the theory was derived, but suggests that some of the conclusions apply more generally. 
In this context, as the simplest estimation for the pairing strength, the BCS equation for $T_\mathrm{c}$ indicates a coupling strength of $\lambda_{\textrm{BCS}}$ $\approx$ 0.1 \cite{supplement}.

%
%

\section*{Discussion}
Taken together, these results reveal an order of magnitude discrepancy between the strength of the $e$-ph coupling that modifies the normal state properties ($\lambda$) and the superconducting pairing strength ($\lambda_{\textrm{BCS}}$), placing SrTiO$_3$ in an unusual regime (Fig. 5D). 
For conventional $e$-ph mediated superconductors, the $e$-ph coupling that renormalizes the band structure is in close quantitative correspondence to the superconducting pairing strength needed to fully account for the superconducting $T_\mathrm{c}$ within ME \cite{McMillan:1965,Allen:1987,Carbotte:1990,Schackert:2015}.  This one-to-one correlation is shown in Fig. 5D, and it should be noted that remarkably few (conventional or unconventional) superconductors lie in the extreme weak-coupling regime.  Considering the strong $e$-ph coupling, occupation of multiple $t_{2g}$ bands, proximity to a ferroelectric quantum critical point, and low carrier density, it would be natural to expect SrTiO$_3$ to exhibit deviations from the BCS limit. Figure 5D shows that while the normal state properties certainly exhibit strong phonon renormalizations, the superconducting state appears conventional and captured by a weak coupling theory. 

In light of this result, it is reasonable to ask whether or not LO4 contributes to pairing and, if so, whether the discrepancy between $\lambda$ and $\lambda_{\textrm{BCS}}$ could be explained by invoking the repulsive Coulomb interaction. 
A conventional  Coulomb pseudo-potential ($\mu$*) approach (questionable in the dilute limit) would require a large $\mu$* ($>$ 0.5), far in excess of canonical values ($< 0.25$) and that varies dramatically with density. 
This, however, is unphysical for SrTiO$_3$ considering the non-interacting behavior found in the dilute limit \cite{Kozuka:2008,Lin:2014}, where the highly polarizable lattice is very effective in dynamically screening the electrons. 
We can infer that the LO4 mode is not effective in mediating an attractive pairing potential, consistent with proposals for other pairing mechanisms which consider exchanging the available low energy modes (i.e. acoustic modes, the TO soft-mode, plasmons, or quantum critical fluctuations) \cite{Appel:1969,Ruhman:2016,Edge:2015a, Klimin:2017}. 

While we cannot distinguish between these possible pairing mechanisms, the results presented here offer further perspective.
A perturbative treatment of the phonon interactions in semiconductors is typically a reasonable approximation, and a natural consequence of the high-energy LO modes is to simply screen both the Coulomb repulsion and the pairing vertex, promoting superconductivity at lower carrier densities than originally considered by BCS. 
However, in most superconducting semiconductors the adiabatic (Migdal) ratio ($\hbar\omega_{ph}/E_\mathrm{F}$) is close to but typically less than 1 \cite{Bustarret:2015}.
It appears essential that in SrTiO$_3$ almost all of the superconducting dome corresponds to $\hbar\omega_\textrm{{LO4}}/E_\mathrm{F} > 1$, and unity ratio corresponds to the loss of superconductivity in the overdoped regime.
Therefore, while large coupling to high energy modes leads to strongly dressed quasiparticles, this interaction does not lead to a comparable contribution to pairing, leaving low-energy phonons (or other bosonic excitations) to give rise to superconductivity out of these polarons, surprisingly well beyond the Migdal limit. 
A comprehensive theory of superconductivity here should realize a weak-coupling superconducting state out of strongly dressed quasiparticles over a wide range of coupling strength and adiabatic parameter.

\section*{Methods}
Nb-doped SrTiO$_3$ films were deposited on TiO$_2$ terminated SrTiO$_3$(001) substrates by pulsed laser deposition as described elsewhere \cite{Kozuka:2010b}. 
Subsequently, 0 - 4 unit cells (u.c.) single crystal LaAlO$_3$ epitaxial layers were grown at $T$ = 650 $^\circ$C with $P$ = $1\times10^{-6}$ torr of O$_2$ with a fluence of 0.43 J/cm$^2$. During growth, the LaAlO$_3$ thickness was monitored by reflection high-energy electron diffraction (RHEED) intensity oscillations. Except where indicated, 3 u.c. thick LaAlO$_3$ tunneling barriers were used throughout this study.
The oxide heterostructure was then post annealed at $T$ = 400 $^\circ$C in 0.4 atm of O$_2$ for 45 min. Atomic force microscopy (AFM) measurements identified that the resulting surface was atomically flat over large areas ($>$ 50 x 50 $\mu$m$^2$) with step-and-terrace morphology.  Nb-doped SrTiO$_3$ films grown in this regime exhibit bulk-like mobility values and full carrier activation \cite{Kozuka:2010b}.
The samples were transferred $ex$ $situ$ to an adjacent chamber for the deposition of metallic electrodes at room temperature through a shadow mask. Just prior to deposition, the films were pre-annealed at 500 $^\circ$C in $1\times10^{-5}$ torr of O$_2$ to remove adsorbates \cite{Inoue:2015}. 
For the superconducting gap measurements, we employed Ag electrodes contacted by Au wire and Ag paint. Ohmic contacts were made to the Nb-doped SrTiO$_3$ film by wire bonding with Al wire. Details regarding the electronic measurements, data analysis, and DOS calculations are provided in the Supplemental Information \cite{supplement}.

\section*{Acknowledgements}
We acknowledge S. A. Kivelson, P. A. Lee, P. B. Littlewood, D. J. Scalapino, A. V. Maharaj, A. Edelman, A. J. Millis, L. Rademaker, Z.-X. Shen, and Y. Wang for useful discussions.  AGS and HI contributed equally.
This work was supported by the Department of Energy, Office of Basic Energy Sciences, Division of Materials Sciences and Engineering, under Contract No. DE-AC02-76SF00515; FAME, one of six centers of STARnet, a Semiconductor Research Corporation program sponsored by MARCO and DARPA (development of polar tunnel barriers); and the Gordon and Betty Moore Foundation's EPiQS Initiative through Grant GBMF4415 (dilution refrigerator experiments). 
S. J. acknowledges support from the University of Tennessee's Science Alliance Joint Directed Research and Development program, a collaboration with Oak Ridge National Laboratory.

\section*{Author Contributions}
AGS and HI prepared the samples, performed the experiments, and analyzed the tunneling data. SJ developed and performed the theoretical calculations. TAM and YH assisted with measurements and sample growth. SJ, TPD, and SR provided theoretical insight and analysis. AGS, HI, and HYH conceived and designed the experiment. All authors contributed to the discussion and writing of the manuscript. AGS and HI contributed equally to this work.


\end{document}